\documentclass[a4paper,11pt]{article}

\usepackage{jinstpub} 

\usepackage{lineno}
\usepackage{subfig}

\title{\boldmath First results from a multiplexed and massive instrument with sub-electron noise Skipper-CCDs}

\author[a,b,1]{F. Chierchie,\note{Corresponding author.}}
\author[c,b,d]{C.R. Chavez,}
\author[e]{M. Sofo Haro,}
\author[c]{G. Fernandez Moroni,}
\author[c,f]{B.A. Cervantes-Vergara,}
\author[g,c]{S. Perez,}
\author[c]{J. Estrada,}
\author[c]{J. Tiffenberg,}
\author[c]{S. Uemura}
\author[c]{ and A. Botti,}

\affiliation[a]{Instituto de Investigaciones en Ingeniería Eléctrica “Alfredo C. Desages” CONICET,\\Bahía Blanca, Argentina}
\affiliation[b]{Universidad Nacional del Sur (UNS), \\ Bahía Blanca, Argentina}
\affiliation[c]{Fermi National Accelerator Laboratory (FERMILAB),\\Batavia, USA}
\affiliation[d]{Facultad de Ingenier\'{i}a - Universidad Nacional de Asunci\'{o}n, \\ Asunci\'{o}n, Paraguay.}
\affiliation[e]{Universidad Nacional de Córdoba, Instituto de Física Enrique Gaviola (CONICET) and \\Reactor Nuclear RA0 (CNEA), Córdoba, Argentina}
\affiliation[f]{Universidad Nacional Aut\'{o}noma de M\'{e}xico, Ciudad de M\'{e}xico, M\'{e}xico}
\affiliation[g]{Universidad de Buenos Aires, Buenos Aires, Argentina}
 
\emailAdd{fernando.chierchie@uns.edu.ar}

\abstract{We present a new instrument composed of a large number of sub-electron noise Skipper-CCDs operated with a two stage analog multiplexed readout scheme suitable for scaling to thousands of channels. New, thick, $1.35$ Mpix sensors, from a new foundry, are glued into a Multi-Chip Module (MCM) printed circuit board on a ceramic substrate which has 16 sensors each. The instrument, that can hold up-to 16 MCMs, a total of 256 Skipper-CCD sensors (called a Super-Module with $\approx 130$ grams of active mass and $346$ Mpix), is part of the R$\&$D effort of the OSCURA experiment which will have $\approx 94$ super-modules. 
Experimental results with $10$ MCMs and $160$ Skipper-CCDs sensors are presented in this paper. This is already the largest ever built instrument with single electron sensitivity CCDs using nondestructive readout, both, in terms of active mass and number of channels.}


\begin{document}
\maketitle
\flushbottom

\section{Introduction}
\label{sec:intro}

The Skipper Charge-Coupled Device (Skipper-CCD) with nondestructive output readout stage can achieve single electron sensitivity allowing to count the exact number of electrons deposited in each pixel \cite{skipper_2012, Tiffenberg:2017aac}. This feature has driven its use in many experiments with various different objectives, from dark matter searches \cite{SENSEI2020}, reactor neutrino experiments \cite{Nasteva_2021}, measurement of properties of silicon \cite{RODRIGUES2021Fano, botti2022constraints}, spectrograph instruments \cite{villalpando2022design} and is now being considered by NASA as a detector for a future space telescope for habitable exoplanets searches \cite{Rauscher2022NASASkipper}.

To increase the sensitivity of the experiments the active mass, and therefore the number of sensors, should be increased significantly. There are instruments with a large number of standard scientific CCDs (noise above one electron). For example, the Dark Energy Camera (DECam) for the Dark Energy Survey has $496$ Mpix using $62$, $2$k$\times 4$k CCDs ($15$ $\mu$m pixels, $250$ $\mu$m thick) with $124$ video channels (additionally $12$ $2$k$\times 2$k CCDs are used for guiding and focus) \cite{Estrada2010DeCAM}, given a total mass of $60$ grams. The CONNIE experiment \cite{CONNIE2019} instrumented $224$ Mpix using $14$ $4$k$\times 4$k standard CCDs ($15$ $\mu$m pixels, $675$ $\mu$m thick) with $28$ video channels, given a total mass of $80$ grams for research in nuclear reactor neutrinos. The DAMIC experiment \cite{Damic2020} instrumented $119$ Mpix using $7$, $4$k$\times 4$k standard CCDs ($15$ $\mu$m pixels, $675$ $\mu$m thick), given a total mass of $40$ grams for dark matter (DM) search. The world's largest camera for astronomy, the LSST (Legacy Survey of Space and Time) camera \cite{2006LSSToverview, 2009lsstCameraElectronics}, currently under construction, has $3.2$ Gpixels using 189 $4$k$\times 4$k CCDs ($10$ $\mu$m pixels, $100$ $\mu$m thick \cite{LSST2019physical}) with a total of $3024$ video channels, $70$ grams of active mass and a performance goal of 5 electrons of noise \cite{2006LSSToverview, 2009lsstCameraElectronics}. 

All the aforementioned instruments use standard scientific CCDs with noise in the range of a few electrons and without the nondestructive readout characteristic. There are also experiments that will reach a large mass of single electron resolution Skipper-CCDs when fully instrumented, such as SENSEI ($100$ grams) \cite{tiffenberg2019sensei} and projected instruments such as DAMIC-M ($1$ kg) \cite{castello2020damicM}.  The OSCURA experiment \cite{aguilar2022oscura} will be the largest instrument using Skipper-CCDs: $10$ kg, $\approx24000$ channels, and $\approx28$ Gpix for DM search. As part of this effort, we present in this paper the largest ever build instrument with single electron sensitivity Skipper-CCDs using nondestructive readout, both, in terms of active mass and number of channels. A pile-up technique \cite{haro2021pile-up} with analog multiplexed \cite{Chavez2022} frontend and new prototype Skipper-CCDs from a new foundry are used for this instrument \cite{cervantes2022skipperMicrochip}.

The organization of this paper is as follows: in Section \ref{sec:InstrumentDescription} the sensors, packaging and electronics of the instrument are described. In Section \ref{sec:Measurments} experimental results are presented performing an individual calibration using the single electron peak and showing subelectron noise operation of the array. Also, results of the first energy spectrum of the interactions measured by combining the data of the best sensors in the array, with an exposure of $0.06$ kg$\times$day, are presented. Finally a summary is presented in Section \ref{sec:summary}.

\section{Description of the instrument}
\label{sec:InstrumentDescription}

\begin{figure}
    \centering
    \includegraphics[width=0.97\textwidth]{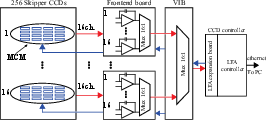}
    \caption{Block diagram of the multiplexed architecture capable of controlling and reading up to $256$ Skipper-CCDs. Red lines indicate the path of video signals and blue lines illustrate the CCDs clock signals, bias voltages and frontend control signals.}
    \label{fig:esquema256}
\end{figure}

In this section we present the main components of the instrument which include: a new sensor package that holds and routes the signals for $16$ Skipper-CCDs called Multi-Chip Module (MCM), the analog multiplexed frontend electronics, a vacuum interface board (VIB), which has a second multiplexing stage, and the readout controller with expanded capabilities.

\begin{figure}
    \centering
    \includegraphics[width=0.97\textwidth]{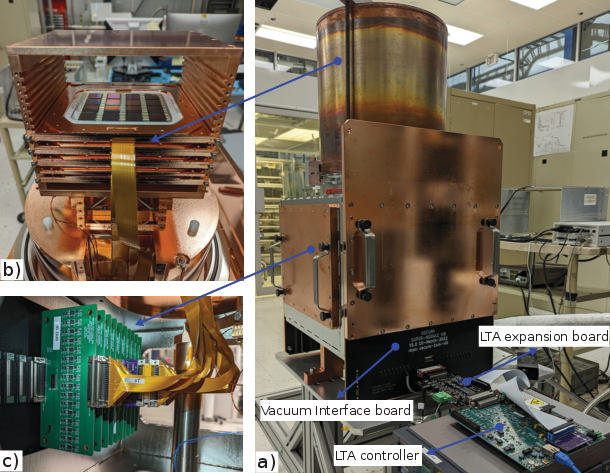}
    \caption{a) Photograph of the exterior of the instrument: cylinder in the top holds the sensors, box in the bottom houses the frontend electronics; b) Copper box and copper trays holding the sensors and flex cables; c) frontend electronics connected to the vacuum interface board. }
    \label{fig:setup256}
\end{figure}

A block diagram of the system can be seen in Fig.~\ref{fig:esquema256} and photographs of different parts of the instruments are shown in Fig.~\ref{fig:setup256}. The vacuum vessel (Fig.~\ref{fig:setup256}a) has two main volumes, a cylinder at the top which holds a copper box with copper trays to hold $16$ modules with $16$ sensors each (Fig.~\ref{fig:setup256}b) and left box in Fig.~\ref{fig:esquema256}), and a box at the bottom where the frontend electronics (Fig.~\ref{fig:setup256}c)) is connected to the vacuum interface board (VIB). Outside the vessel, the Low Threshold Acquisition Controller (LTA) \cite{cancelo2021low} with an expansion board, developed for this instrument, is used to generate all the clocks and reference voltage required to operate the array of Skipper-CCDs and also to digitize the pixels values. A cryocooler is used to operate the sensors at temperatures between $130$ and $160$ Kelvin cooling down the copper box and trays. As shown in the block diagram of Fig.~\ref{fig:esquema256}, all the frontend boards share the digital control signals and the MCMs share the CCDs clock signals and bias voltages (blue lines). Each of the 16 MCMs has it own frontend board that includes the first multiplexing stage. The VIB has the second multiplexing stage and is connected from the outside to the LTA expansion board, video lines are shown with red lines.

\subsection{New sensors and Multi-Chip Module Package}

Since the first development of the Skipper-CCD operating with deep sub-electron noise \cite{skipper_2012, Tiffenberg:2017aac}, the sensors were fabricated at Teledyne DALSA. Due to plans for stopping the production line required for thick fully depleted CCDs, the OSCURA project has fabricated the first run of new prototype sensors (designed at Lawrence Berkeley National Laboratory, LBNL) at Microchip Technology Inc in $200$ mm diameter wafers, for details of the individual performance of these new sensors see \cite{cervantes2022skipperMicrochip}. The Skipper-CCDs adopted for OSCURA are thick ($675$ to $725$ $\mu$m) sensors with $1278$ rows and $1058$ columns given a total of $1.35$ Mpix per sensor. Each sensor is readout using one out of the four available amplifiers. The small sensors area guarantees a high yield, larger sensors are prone to fabrications defects, and also allows a higher readout speed (smaller number of pixels to be readout per amplifier), this is required to reduce the dark-current rate which is proportional to the exposure and readout times.

\begin{figure}
    \centering
    \includegraphics[width=0.75\textwidth]{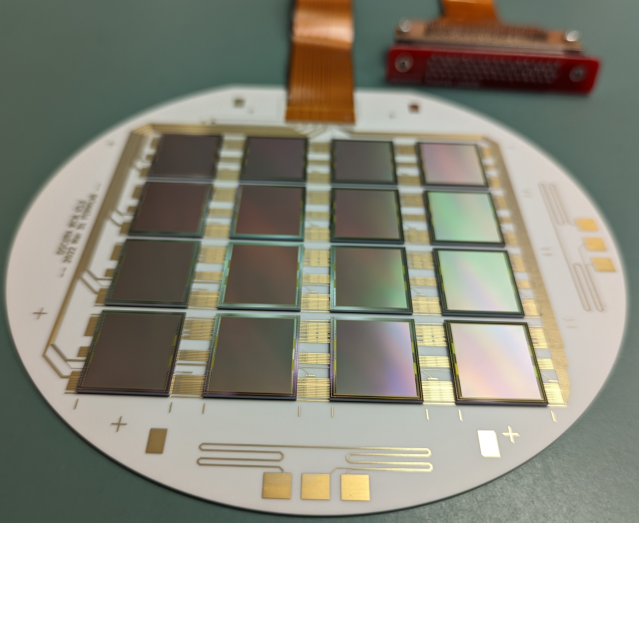}
    \caption{Photograph of a Multi-Chip Module (MCM): single layer ceramic substrate with 16 glued single-electron resolution Skipper-CCDs.}
    \label{fig:Ceramic_mcm}
\end{figure}

Sixteen sensors are assembled in a single module called Multi-Chip Module (MCM). The printed circuit board is a single layer in a ceramic substrate (96$\%$ Al$_{2}$O$_{3}$) of $0.635$ mm thick. The traces are made of copper $34.8$ $\mu$m thick with a surface finishing of ENEPIG. This ceramic offers electrical isolation and a good thermal conductivity for cooling down the sensor, which are glued on top of the copper traces using an epoxy. The flex cable is also glued to the ceramic and all connections, to the flex, to the sensors and between traces are made by wire-bonds. Figure \ref{fig:Ceramic_mcm} shows a photograph of the MCM where the traces, 16 sensors, flex cable and some test structures are observed. A $68$ pin connector is used to attach the module to the frontend electronics. For the OSCURA experiment research and development efforts are being made to replace the ceramic substrate with a silicon substrate which will be of the same quality as the one used to produce the sensors. This will reduce the radioactivity and lower any possible background signal produced by the contamination in the substrate material, as will be shown in the experimental measurements.

\subsection{Electronics}

The MCM has no active electronics other than the output transistors in the CCD sensors. The flex cables are routed to the bottom of the vessel (see Fig.~\ref{fig:setup256}), and connected to the frontend electronics board, one for each MCM as schematically shown in  Fig.~\ref{fig:esquema256}. This board has sixteen analog channels, one per sensor, to compute the pixel values and an analog multiplexer to read one channel at a time. The analog processing chain of each channel consist of:  a preamplifier (gain $5$), an analog integrator circuit that implements a novel pile-up technique, which adds up independent measurements of the same pixel charge (skipper samples) into the integrator's capacitor \cite{haro2021pile-up}. The pile-up technique highly reduces the data rate because only the final averaged pixel value is readout. The final stage is a sample and hold circuit, that could be used to store the pixel value before ADC conversion. Details of the operation, capabilities and advantages of using this frontend circuit were reported in \cite{Chavez2022}. 

The other side of the frontend electronics is connected to a new-designed Vacuum Interface Board (VIB). In the vacuum side, this VIB has sixteen 68-pin connectors for all the required frontend boards (see Fig.~\ref{fig:setup256}c)). On the outside, the VIB has a second stage of analog multiplexing and a $50$ pin connector which is enough to route all the control signals for the frontend, two multiplexing stages, CCD clock signals, CCD reference voltages and a single multiplexed video line. 

The LTA \cite{cancelo2021low}, which has $4$ fully digital video channels, and  was originally designed to read a single four channel CCD, was used for readout and control of the sensors and frontend. Only a single channel and ADC of the LTA is necessary for this instrument. An expansion board was designed and firmware modifications were made to be able to sequence additional clock signals required to control the analog switches of the frontend and the control signals of the two stage multiplexing. With these upgrades, the full instrument could be controlled using a single LTA. The expansion board also includes high current buffer for the vertical transfer clocks of the CCDs, in case they are required, and a video offset correction circuit with a low noise reference voltage that could be used to center the output into the ADC converter range. It also allows the use of external power supplies for any reference voltage that could exceed the output current capabilities of the LTA.

A clock sequencer to control the 256-ch system was programmed. A single multiplexed image is generated by the software in each acquisition and a post processing script is used to demultiplex and generate one 16-channel image per MCM.



\section{Measurements and experimental results}
\label{sec:Measurments}

The instrument is partially loaded with $10$ MCMs, i.e. a total of $160$ sensors, this is already the largest array of single electron sensitivity Skipper-CCDs ever built with $84$ grams of active sensors. The total power delivered to the $10$ frontend boards and LTA expansion board is $31.8$ Watts. An external power supply for the polarization of the output transistors of the CCDs was used due to the large amount of amplifiers ($\approx 0.63 $ mA per amplifier), which exceeds the capability of the LTA controller. All other clocks and bias voltages, including the $70$ V substrate voltage to fully deplete the sensors, are provided by the LTA.

\begin{figure}
    \centering
    \includegraphics[width=0.99\textwidth]{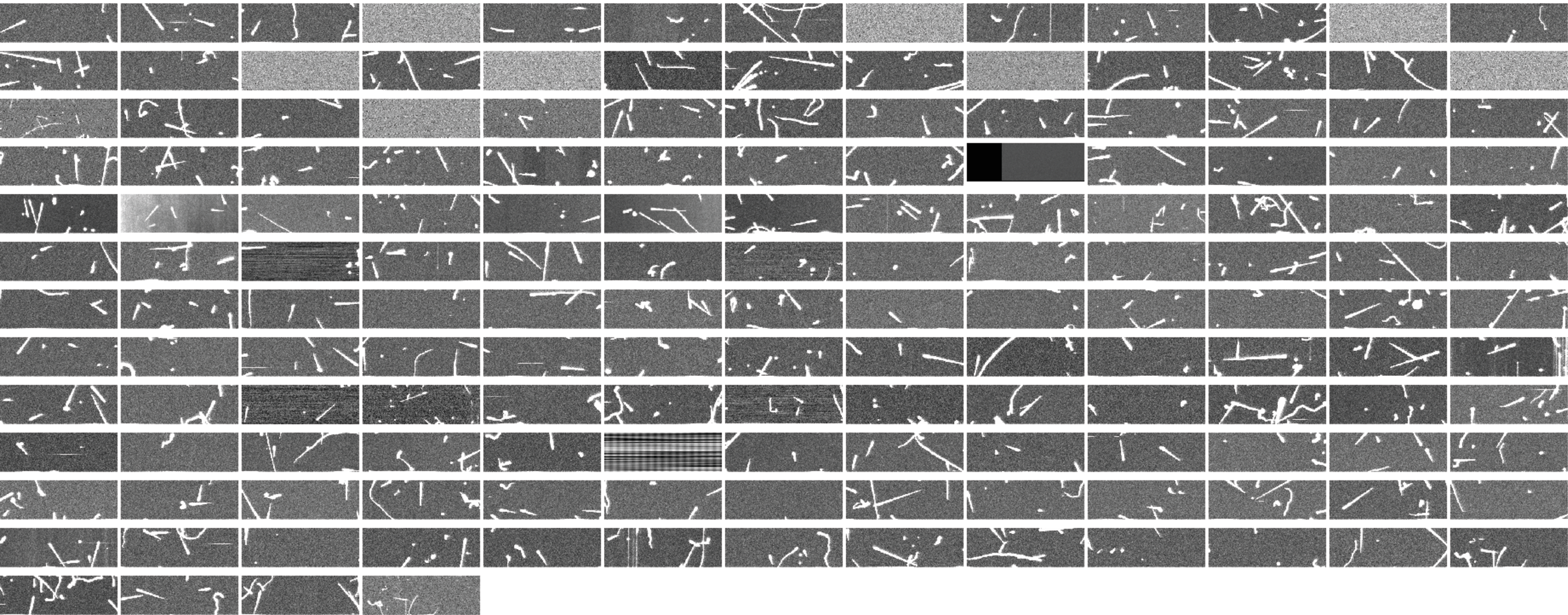}
    \caption{An image showing sub-images with cosmic rays interactions of all the $160$ sensors of the array.}
    \label{fig:allChIm}
\end{figure}

Figure \ref{fig:allChIm} shows an image with $160$ sub-images showing a fraction of each sensor active area after a few minutes of exposure. Most of the channels show tracks produced by muons and electrons from cosmic rays. A few malfunctioning sensors were disconnected (by removing the wire-bonds). There was no preselection or pretesting of the sensors before gluing them into the MCMs, despite of this a large yield, above $90\%$ of the sensors worked and presented a good performance in terms of the noise. 

\begin{figure}
    \centering
    \includegraphics[width=0.99\textwidth]{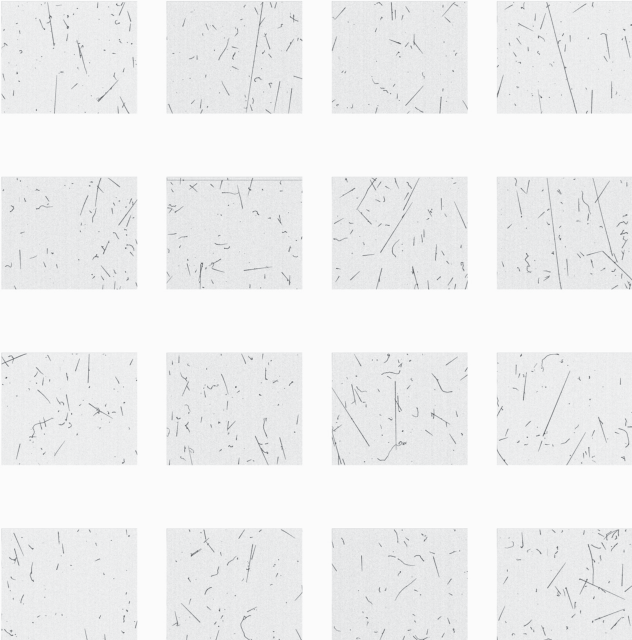}
    \caption{Full image of a single MCM with 16 sensors. Each sub-image corresponds to a $1.35$ Mpix sensor and are separated by the space in the physical arrange. Muons crossing more than one sensor can be appreciated. The color scale in this image was inverted compared to Fig.~\ref{fig:allChIm}. }
    \label{fig:focalplaneImage}
\end{figure}

Figure \ref{fig:focalplaneImage} shows a detailed view of a full image acquired by one MCM. The 16 sub-images are arranged in the positions and with the separations that the sensors have in the physical array (see Fig.~\ref{fig:Ceramic_mcm}). Cosmic rays interactions are seen in all the channels, some of the tracks produced by muons (straight lines) cross more than one sensor, for example as seen in the images produced by the sensors of the first and second row of the last column.

\subsection{Single electron operation and calibration}

The gain or conversion factor between the charge deposited in the pixel (in units of $e^{-}$) and the number of digital to analog converter units (ADU) is given by \cite{Chavez2022}
\begin{equation}
G[ADU/e-] = k  A  N  \left(\frac{t_i}{R\times C}\right) S 
\label{eq:gain}
\end{equation}
where: $S$ $[\mu V/e^-]$ is the sensitivity of the sense node in the CCDs (usually in the range between $1.2$ and $2.5$ $[\mu V/e^-]$), $R=2000$ $\Omega$ and $C=18$ nF are the resistor and capacitors of the integrator circuit in the frontend electronics, $t_i$ is the integration interval during the pedestal and signal periods of the video signal ($t_i=10$ $\mu$s  was used for the experiments), $N$ is the number of times the charge in each pixel is measured (Skipper samples) using the pile-up technique, $A=16.5$ $[V/V]$ is the total amplification gain ($5$ in the preamplifier and $3.3$ in the LTA) and $k=0.128$  $[ADU/\mu V]$ is the ADC conversion factor. For example, replacing in (\ref{eq:gain}) using a sensitivity of $1.7$ $[\mu V/e^-]$ and dividing by $N$ gives the nominal normalized gain of $G/N \approx 1$ $[ADU/(N\times e^-)]$.
\begin{figure}
    \centering
    \includegraphics[width=0.97\textwidth]{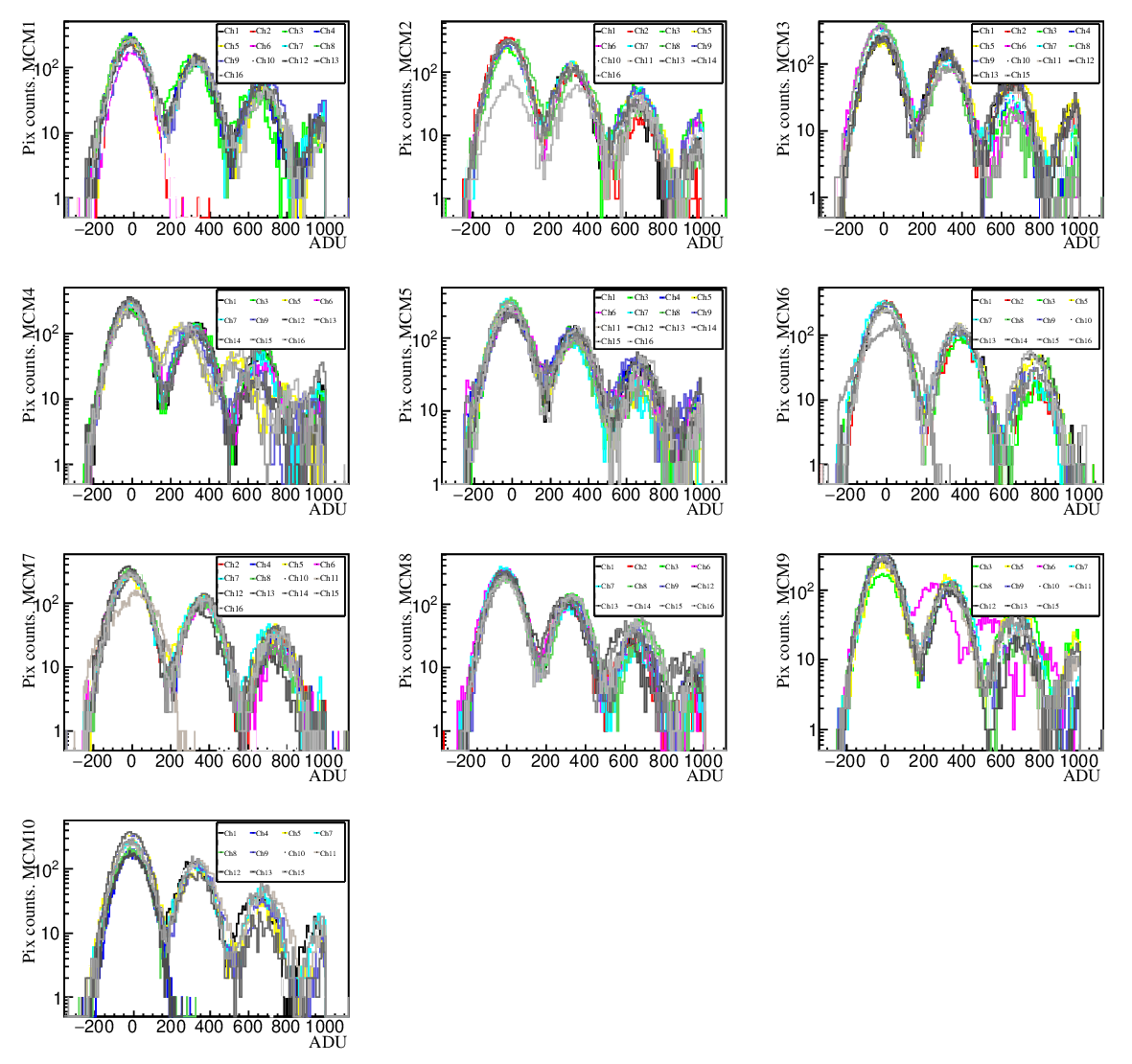}
    \caption{Histograms of pixels for each of the MCMs and channels for an acquisition with $N=400$ samples per pixel and pixel integration time of $t_i=10$ $\mu$s.}
    \label{fig:peaks}
\end{figure}

An image was acquired with $N=400$ and $t_i=10$ $\mu$s and a histogram of a subset of the pixels in the active area was computed for each of the MCMs and for each of the channels. The results are shown in Fig.~\ref{fig:peaks}, with the labels indicating the channels that presented a noise below $0.22$ $e^-$, the average noise of the channels is around $0.175$ $e^-$. The peaks produced by the charge quantization and the sub-electron noise of the sensors are clearly observed. These measurements shows the capability of the instrument for achieving the same noise performance that was previously obtained for a single sensor (or a few of them) but with hundreds of Skipper-CCDs.
The gain of each channel was calibrated by fitting a Gaussian to the $0$ $e^-$ electron peak and $1$ $e^-$ which are the two peaks with higher amplitude. These gains $G[ADU/e^-]$ for each channel are used to calibrate all other measurements in this work.

\begin{figure}
    \centering
    \includegraphics[width=0.97\textwidth]{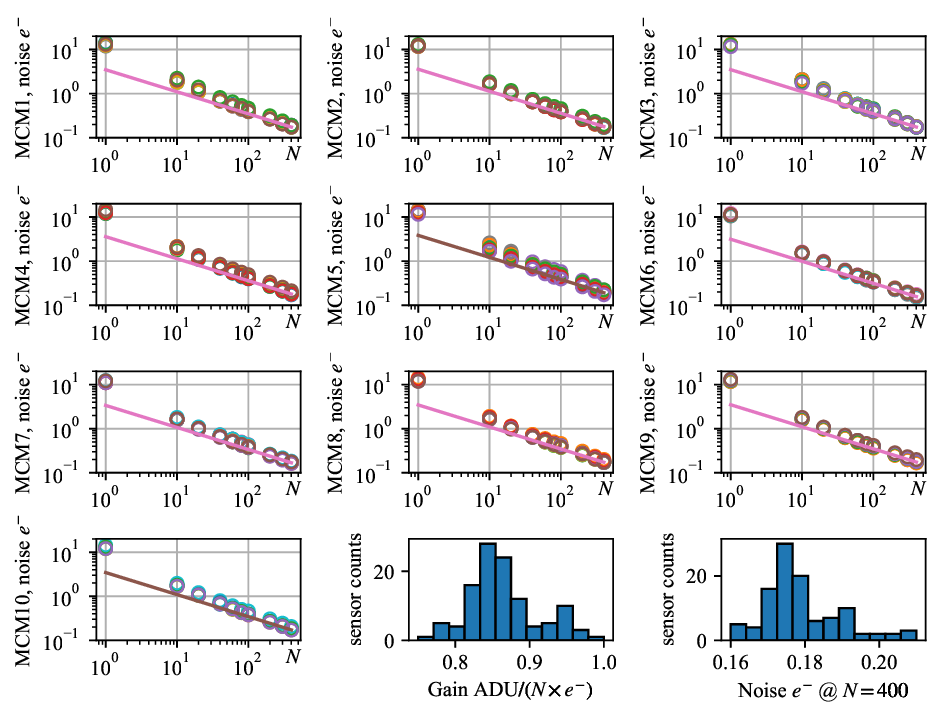}
    \caption{Noise in $e^-$ as a function of the number of samples per pixel $N$ for all the MCMs and channels. The two histograms at the right bottom show the gain and noise computed for $N=400$.}
    \label{fig:NoiseScanAndGain}
\end{figure}

Figure \ref{fig:NoiseScanAndGain} shows measurements of the noise in  $e^-$ computed for $10$ different values of $N$ in the range between $N=1$ and $N=400$ for all the MCMs. A desirable homogeneous response of all the channels is observed, since most of the circular markers are superimposed. For each MCM the $1/\sqrt{N}$ expected noise reduction for independent samples averaging, which is seen as a straight line in a log-log plot is shown. This rate was computed by taking as starting point the noise at $N=400$ for the channel with the closest performance to the median noise. For $N>20$ the noise performance follows the expected reduction rate and circles are over the line. As explained in \cite{haro2021pile-up}, the noise for $N=1$ is not dominated by the CCD performance but by the analog readout electronics because the frontend is designed for operation with $N\gg 1$. The two histograms shown in the last row of Figure \ref{fig:NoiseScanAndGain} were computed with the measured gains $G/N$ and noises in $e^-$, for $N=400$ and for all the sensors.

\subsection{Energy spectrum}


The noise performance as a function of $N$ and the ability of the system for single electron counting in more than a hundred sensors was experimentally demonstrated in the previous section. In this section the number of sample per pixel was fixed at $N=20$, which according to the results in Fig.~\ref{fig:NoiseScanAndGain} results in a noise of around  1 $e^-$. In this mode of operation the analog frontend electronics can achieve a wide dynamic range allowing to calculate an spectrum in a higher range of energies. A total of $100$ acquisitions were taken ($100$ images per channel) for these measurements.

To compute the spectra, several post-processing steps are applied to the raw images: 1) baseline subtraction based on an overscan region of the image is done line by line, 2) master bias subtraction of each channel is performed by computing the median image over the $100$ acquisitions and subtracting the result from all the images and 3) a clusterization algorithm is applied to add up the energy of each event spread out in more than one pixel. As a result, a catalog of the events with information of its energy, position and additional geometric metrics, such as variance of the size in the two dimensions of the images is build. The calibration gain for each channel, presented in the previous section, was used and an electron-hole pair creation energy of $3.75$ eV \cite{RODRIGUES2021Fano} was applied to calibrate in units of electron-volts. No shielding was used for the experiment so a high background rate is expected.

\begin{figure}
    \centering
    \subfloat[]{\includegraphics[width=0.99\textwidth]{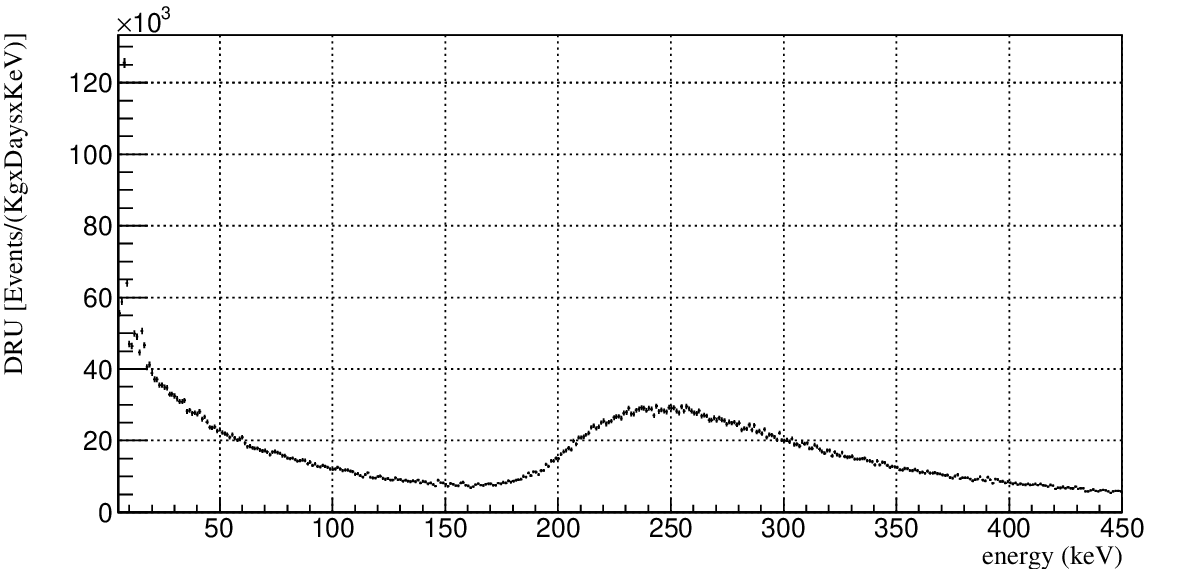} \label{fig:spectrumMuons}}
    \hfill
    \subfloat[]{\includegraphics[width=0.99\textwidth]{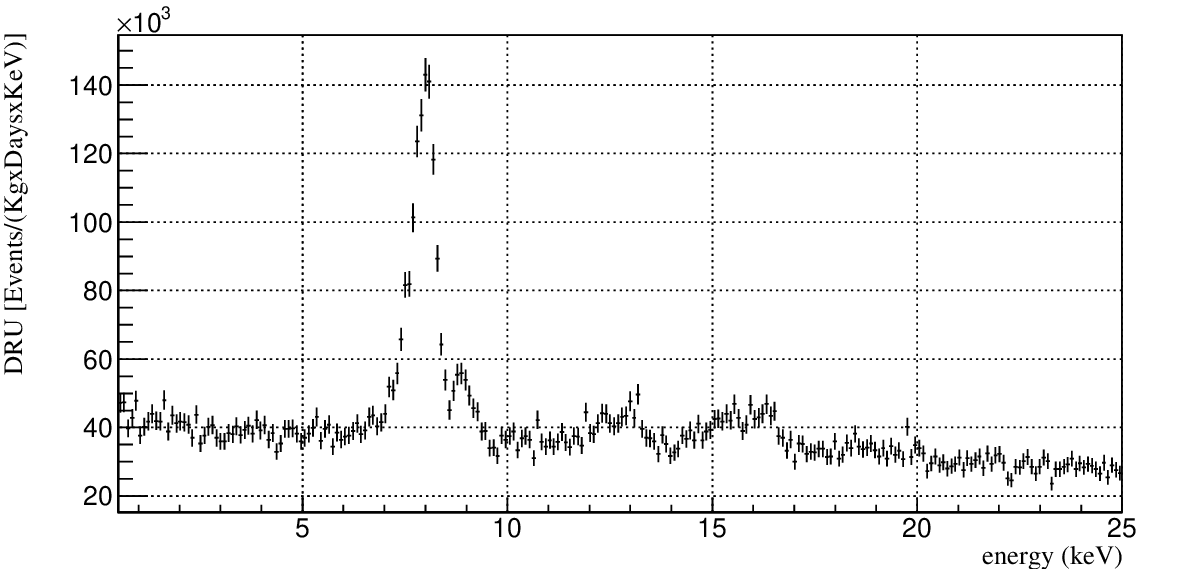} \label{fig:spectrumCu} }
    \caption{Spectra in two different ranges of energy obtained for mass$\times$time of $0.06$ kg$\times$day. Spectra are scaled by the channel selection cut and active area region used. In \ref{fig:spectrumMuons} no additional selection cuts were applied; in \ref{fig:spectrumCu} geometric selection cuts were applied as described in the text, the spectrum is not scaled by the efficiency of these cuts.}
    \label{fig:spectrum}
\end{figure}

Two ranges of energy were used as shown in Fig.~\ref{fig:spectrum}. To compute these spectra a selection of the MCMs and channels with the best performance was made. The active region of the sensors was used excluding a $50$ pixels border all around the edges of the active area. After the quality channel selection and pixel area the total active mass used is $0.046$ kg and the average exposure/readout time of one image is $0.013$ days resulting in a mass$\times$time of $0.06$ kg$\times$day for the $100$ measurements.


Figure \ref{fig:spectrumMuons} shows the resulting energy spectrum between $5$ keV and $450$ keV. The most noticeable feature is the bump in the range between $200$ keV and $350$ keV, this is produced by the charge deposited in the pixels mostly by muons interactions. No selection cuts were applied for this spectrum other than the previously mentioned: exclusion of the borders of the active area and channels selection, and the spectrum is scaled by these cuts.

Figure \ref{fig:spectrumCu} shows the spectrum in an energy range between $0.5$ keV and $25$ keV. For this spectrum, additional selection cuts to filter events based on their shape were applied. The standard deviation of the size in the $x$-direction $\sigma_x$ and in the $y$-direction $\sigma_y$ were limited \cite{Moroni2022SkipperAboveGround} to extract single hits taking into account the expected diffusion for those events produced by the thickness of the sensor \cite{haro2020studies}. 
The spectrum in Fig.~\ref{fig:spectrumCu} is scaled by the channel selection cut and active area region used but not by the geometric selection cuts. 

X-ray fluorescence lines produced by the materials surrounding the detectors can be observed in Fig.~\ref{fig:spectrumCu}, the most noticeable are two copper peaks emitted by the cold box and copper trays seen in \ref{fig:setup256}. These are $k_{\alpha 1,2}=8.05,8.03$ keV seen as the single most noticeable peak and the $k_{\beta}=8.91$ keV peak. Despite using the calibration obtained in the previous section at very low charge, with the zero and one electron peak, a good linearity is obtained reflected in the position of these peaks produced by thousands of deposited electrons.

Two additional, more disperse peaks, are observed in the ranges $12-14$ keV and $15-17$ keV. Similar spectral characteristics were reported in the engineering run of the CONNIE experiment \cite{CONNIEaguilar2016engineeringRun} using standard non-Skipper-CCDs and in the DAMIC experiment \cite{settimo2018damic}, specifically in \cite{tiffenberg2013damic} peaks in the ranges $12-14$ keV and $15-17$ keV were identified as produced by uranium contamination in the ceramic aluminum nitride (AlN) substrate that provides mechanical support to the CCDs in those experiments. The circuit on ceramic substrate of the MCMs will be replaced by a circuit on a high purity silicon substrate for the OSCURA experiment to reduce this radioactive contamination. This circuit on silicon is currently under development and will be reported in the future.

\section{Summary}
\label{sec:summary}
The largest instrument built to date using single electron sensitivity Skipper-CCDs with nondestructive readout was presented in this paper. A two level analog multiplexed readout scheme and a charge pile-up frontend that allow to average the multiple measurements of the same charge packet were demonstrated using hundreds of sensors. Experimental results showing the capability for counting single electrons and the flexibility of computing a higher energy spectrum by changing the number of samples per pixel were also shown. This instrument can have direct application in dark matter and neutrino searches and the readout approach is scalable to thousands of sensors representing an important result towards the construction of the $10$ kg OSCURA experiment.

\acknowledgments
The authors would like to thank the SiDet team at Fermilab for the support on the development of this instrument: Michelle J. Jonas for the assembly and wire-bonding of the MCMs, Leland E Scott Jr. for the assembly of the many printed circuits boards required, Eng. Hemanth Kiran Gutti for the mechanical designs and specially to Andrew Lathrop for always being available to solve any issue or building any necessary parts.
Lawrence Berkeley National Laboratory is the developer of the fully depleted CCD and the designer of the Skipper-CCD readout.


\end{document}